\documentclass[aps,prl,twocolumn,superscriptaddress]{revtex4}
\usepackage{amsmath}
\usepackage[colorlinks]{hyperref}
\usepackage{amssymb}
\usepackage{color}
\usepackage{graphicx,amsmath}
\UseRawInputEncoding \input
\hypersetup{colorlinks,citecolor=red,linkcolor=blue,urlcolor=blue}

\begin{document}
\title{Comment on "Stranger than metals"}
\author{V. R. Shaginyan}\email{vrshag@thd.pnpi.spb.ru} \affiliation{Petersburg
Nuclear Physics Institute of NRC "Kurchatov Institute", Gatchina,
188300, Russia}\affiliation{Clark Atlanta University, Atlanta, GA
30314, USA} \author{A. Z. Msezane}\affiliation{Clark Atlanta
University, Atlanta, GA 30314, USA}
\author{G. S. Japaridze}\affiliation{Clark Atlanta
University, Atlanta, GA 30314, USA} \author{M.~V. Zverev}
\affiliation{NRC Kurchatov Institute, Moscow, 123182, Russia}
\affiliation{Moscow Institute of Physics and Technology,
Dolgoprudny, Moscow District 141700, Russia}

\begin{abstract}
P. W. Phillips, N. E. Hussey, P. Abbamonte (Review Article, 8 July
2022, eabh4273) consider heavy fermion (HF) metals and high-$T_c$
superconductors naming them strange metals. They analyze such
features of strange metals as quantum criticality, Planckian
dissipation and recently observed fundamental link between the
high-$T_c$ superconductivity and strange metals, and conclude that
these problems can be possibly resolved within the framework of
theories based on gravity, etc. In this comment we discuss that this claim is
not correct and the successful description of the quantum
criticality, Planckian dissipation and recently observed
fundamental link between the high-$T_c$ superconductivity and
strange metals has been given within the framework of the fermion
condensation theory.
\end{abstract}

\maketitle

Authors of the review \cite{sci22,arx22}, analyzing heavy fermion
(HF) metals and high-$T_c$ superconductors, dubbed them strange
metals, announce "given the immense difficulty in constructing a
theory of strange metals, one might ask why bother?" \cite{sci22}.
We concur with this claim, since the well-known fermion
condensation (FC) theory describes the quantum criticality,
Planckian dissipation, recently observed fundamental link between
the high-$T_c$ superconductivity and strange metals, and the other
thermodynamic, transport and relaxation properties, see e.g.
\cite{ks,vol,physrep,zsp,shagrep,prb2013,prb2016,pccp,prb2020,springer}.
The FC theory is based on the notion of flat band, that was
predicted in 1990 \cite{ks} and now represents exciting and hot
topic in condensed matter physics, see e.g. \cite{catal,bern}. In
the FC theory the quantum critical point is represented by the
topological fermion condensation quantum phase transition (FCQPT)
that increases the Fermi surface dimension by one, creating both
the flat band and the new state of matter
\cite{ks,vol,physrep,shagrep,springer}.
\begin{figure}[!ht]
\begin{center}
\includegraphics [width=0.47\textwidth]{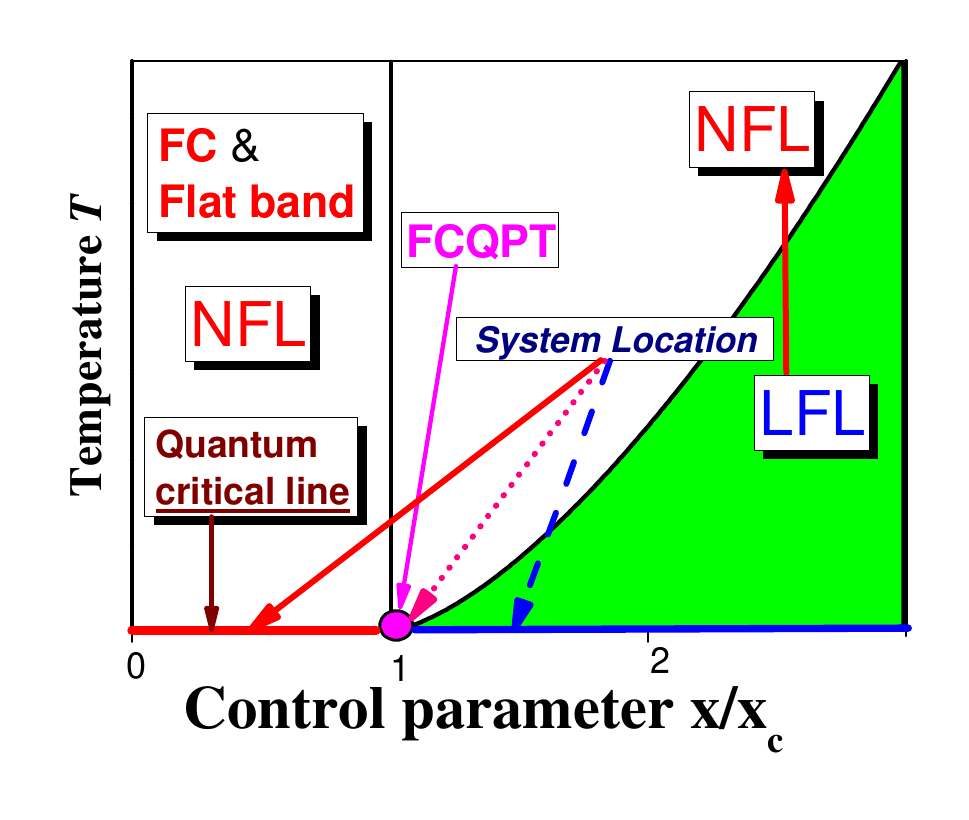}
\end{center}
\caption{Schematic $T - x/x_c$ phase diagram of the system with FC.
The diagram is adapted from \cite{shagrep,prb2016}. The
dimensionless doping $x/x_c$ is the control parameter. The system's
locations are depicted by the arrows. At $x/x_c<1$ the system is
shifted beyond FCQPT, the corresponding location shown by the solid
arrow. At $x/x_c<1$ the system is at its quantum critical line,
possesses a flat band, and at any finite temperature the system
exhibits the NFL, or critical behavior with the resistivity given
by Eq. \eqref{rho}. The dotted line shows the system located at
FCQPT, $x/x_c=1$. The dashed arrow points to the system at
$x/x_{c}>1$ when it is in the LFL state at sufficiently low
temperatures, shown by the shaded area. The vertical arrow shows
the system moving from LFL to NFL at $T$ increasing and the control
parameter fixed.} \label{fig1}
\end{figure}
The schematic phase diagram of the system which is driven to FC
state with a flat band, by varying the dimensionless doping
parameter $x/x_c$ is reported in Fig. \ref{fig1}. The tuning
parameter $x/x_c$ can represent  other parameters besides doping,
such as pressure $P$, $P_c$ being the corresponding critical value
\cite{shagrep,prb2016}. Upon approaching the critical density
$x_{c}$ the system remains in the Landau Fermi liquid (LFL) region
at sufficiently low temperatures as it is shown by the shaded area.
At FCQPT ($x/x_{c}\leq1$) shown by the solid arrow in Fig.
\ref{fig1}, the system is at quantum critical line, and
demonstrates the non-Fermi liquid (NFL) behavior, that is critical
behavior with $\rho(T)$ exhibiting \cite{shagrep,prb2016}
\begin{equation}\label{rho}
\rho(T)= A_1T.
\end{equation}
Such a behavior is ubiquitous for systems with flat bands, see e.g.
behavior exhibited by the HF metal $\rm {\beta-YbAlB_4}$
\cite{springer,prb2016}. At $x/x_{c}\leq1$ the system becomes
superconducting with $T_c\propto g$, provided that the
superconducting coupling constant $g>0$, see e.g.
\cite{ks,physrep,shagrep,springer,bern}.
\begin{figure}[!ht]
\begin{center}
\includegraphics [width=0.47\textwidth]{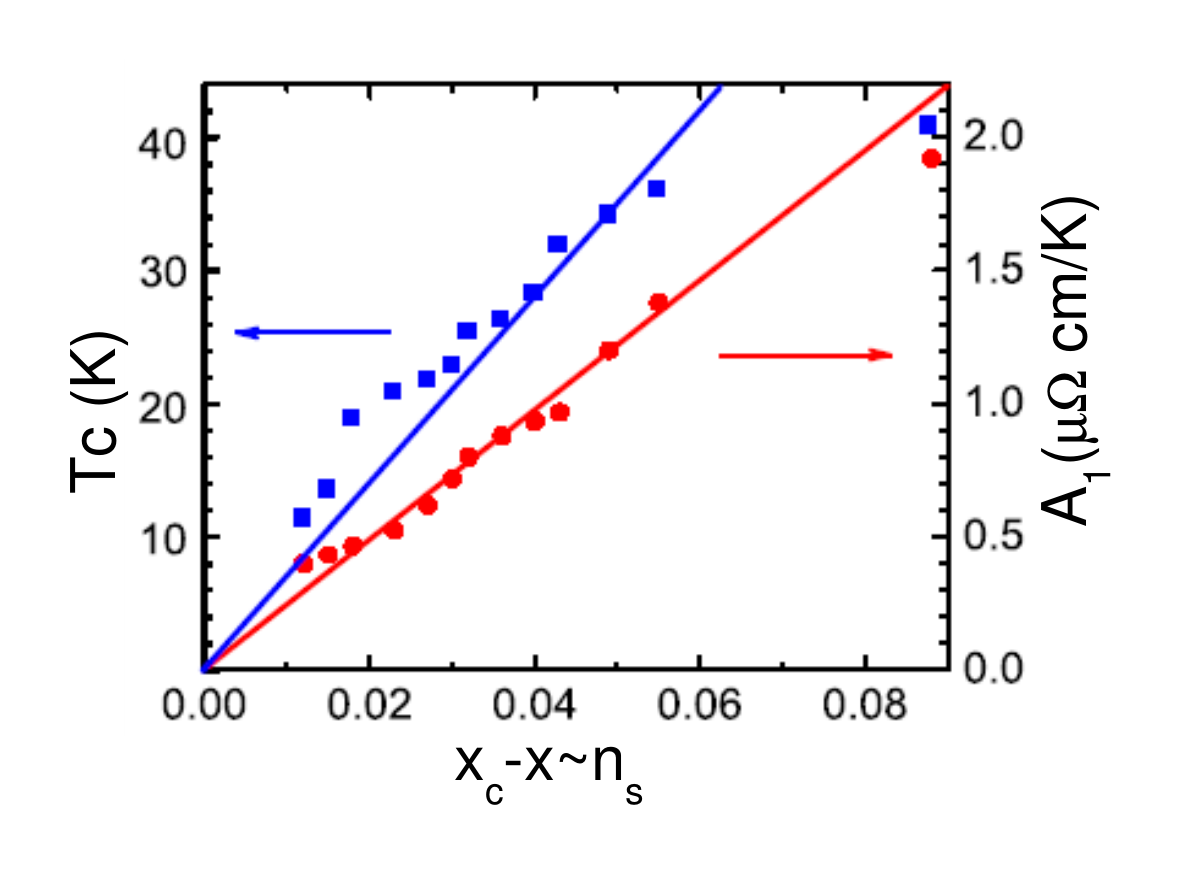}
\end{center}
\caption{Dependence of the coefficient $A_1$ in the resistivity
$\rho(T)$ (red circles, right axis) and the critical temperature
$T_c$ (blue squares, left axis) of overdoped $\rm
La_{2-x}Sr_xCuO_4$ films on the doping x measured from its critical
value $x_c=0.26$ \cite{bosovic}. Red and blue lines show the best
linear fits to the data, which support the conclusion that
$A_1(x)\propto T_c(x)\propto n_s$, indicative of behavior
inconsistent with conventional theories. Adapted from
\cite{prb2020}.} \label{fig2}
\end{figure}
At $x/x_{c}\simeq1$ the NFL state above the critical line is
strongly degenerate. As a consequence, with different phase
transitions emerge, lifting the degeneration. Thus, the NFL state
can be captured by the other states like the superconducting one
\cite{shagrep,prb2016,springer}. The diversity of phase transitions
occurring at low temperatures is one of the most spectacular
features of the physics of HF metals, and is ubiquitous to systems
with flat bands.

\begin{figure} [! ht]
\begin{center}
\includegraphics [width=0.55\textwidth] {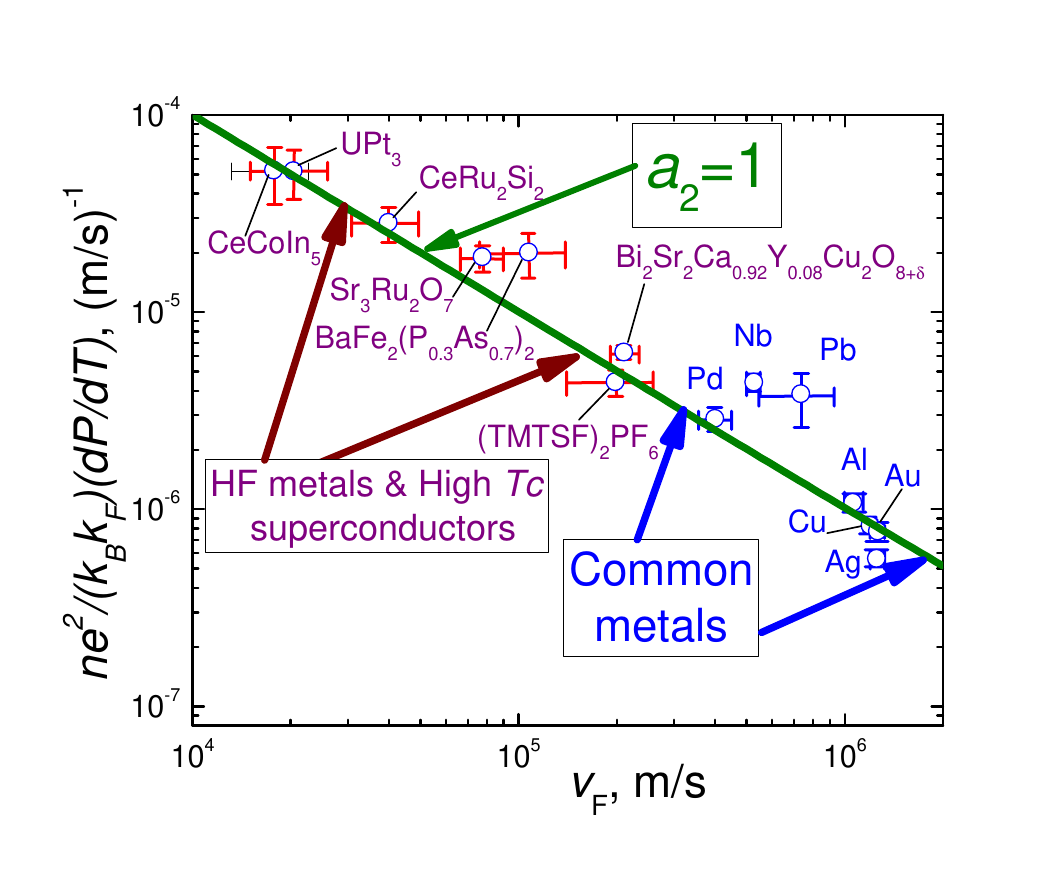}
\end {center}
\caption{Scattering rates of different strongly correlated metals
like HF metals, high-$T_c$ superconductors, organic metals, and
conventional metals \cite{bruin}; adapted from \cite{mdpi2022}. All
these metals exhibit $\rho(T)\propto T$ and their Fermi velocities
$V_F$ vary by two orders of magnitude. The parameter $a_2\simeq 1$
gives the best fit shown by the solid green line. The region
occupied by the common metals is displayed by the two blue arrows,
and the two maroon arrows show the region of strongly correlated
metals} \label{Sc1}
\end{figure}

Let us note that the coefficient $A_1$, see Eq. \eqref{rho}, tracks
the doping dependence of both superfluid density $n_s$ at $T=0$ and
$T_c$ \cite{bosovic}. It is seen from Fig. \ref{fig2} that in
high-temperature superconducting overdoped copper oxides where
$T_c(x)$ terminates at critical doping value $x_c$, the quite
remarkable dependence $A_1(x)\propto T_c(x)\propto n_s(x)$ has been
discovered \cite{bosovic,sci22}. This exciting behavior has been
predicted \cite{zsp}, and explained within the framework of the FC
theory \cite{pccp,springer,prb2020}.

The Planckian dissipation and the linear dependence of the
resistivity, see Eq. \eqref{rho}, are also explained within the
framework of FC theory. It is remarkable that the Planckian
dissipation is demonstrated by both strange metals at their quantum
criticality and conventional metals \cite{bruin}, since the same
physics describes the $\rho(T)\propto T$ dependence of these metals
\cite{prb2013,prb2020,springer}, see Fig. \ref{Sc1}.

In summary, we have shown that the problems outlined in Review
\cite{sci22} as unresolved enigmas of strange metals, have been
explained within the framework of the FC theory. Moreover, there
are a few numbers of important features demonstrated by strange
metals that have been missed in Review \cite{sci22}. Among these
are (see e.g. \cite{mdpi2022}):

The universal $T/B$ scaling behavior of the thermodynamic and
transport properties, including the negative magnetoresistance of
strange metals;

The recent challenging experimental facts regarding the tunneling
differential conductivity $dI/dV=\sigma_d(V)$, as a function of the
applied bias voltage $V$, that are collected under the application
of magnetic field $B$ on the twisted graphene and the archetypical
heavy fermion metals $\rm YbRh_2Si_2$ and $\rm CeCoIn_5$;

The emergence of the asymmetrical part of the tunneling
conductivity (or resistivity)
$\Delta\sigma_d=\sigma_d(V)-\sigma_d(-V)$ as well as that
$\Delta\sigma_d$ vanishes in magnetic fields, as was predicted
\cite{shagrep}.

Transition temperature $T_c$ is proportional to the Fermi velocity
$V_F$, $V_F\propto T_c$, rather than $1/V_F\propto T_c$, as it
takes place in common Bardeen-Cooper-Schrieffer like theories;

Flat bands make $T_c\propto g$, with $g$ being the superconducting
coupling constant;

Thus, these recent outstanding experimental results strongly
suggest that the topological FCQPT is an intrinsic feature of many
strongly correlated Fermi systems, and can be viewed as the
universal agent defining their non-Fermi liquid behavior.  And the
fermion condensation theory is able to explain challenging features
exhibited by strongly correlated Fermi systems, including strange
metals.

\end{document}